\documentclass[multphys,vecphys]{svmult}

\usepackage{makeidx}         
\usepackage{graphicx}        
                             
\usepackage{multicol}        
\usepackage[bottom]{footmisc}

\newcommand{\rmn}{\mathrm}
\newcommand{\M}{{\mathcal M}}
\newcommand{\dd}{\mathrm{d}}
\newcommand{\eps}{\varepsilon}
\newcommand{\CR}{{\rm CR}}
\newcommand{\therm}{{\rm th}}

\makeindex

\begin{document}

\title*{Cosmological structure formation shocks and cosmic rays in hydrodynamical
simulations}
\titlerunning{Cosmological shocks and cosmic rays in hydrodynamical
simulations}
\author{C. Pfrommer\inst{1}, V. Springel\inst{2}, T.A. En{\ss}lin\inst{2}, \and
  M. Jubelgas\inst{2}}

\institute{Canadian Institute for Theoretical Astrophysics, University of Toronto,
  60 St. George Street, Toronto, Ontario, M5S 3H8, Canada \\
\texttt{pfrommer@cita.utoronto.ca}
\and Max-Planck-Institut f\"ur Astrophysik, Karl-Schwarzschild-Stra{\ss}e 1,
  Postfach 1317, 85741 Garching, Germany}

\maketitle

\abstract Cosmological shock waves during structure formation not only play a
decisive role for the thermalization of gas in virializing structures but also
for the acceleration of relativistic cosmic rays (CRs) through diffusive shock
acceleration.  We discuss a novel numerical treatment of the physics of cosmic
rays in combination with a formalism for identifying and measuring the shock
strength on-the-fly during a smoothed particle hydrodynamics simulation.  In
our methodology, the non-thermal CR population is treated
self-consistently in order to assess its dynamical impact on the thermal gas as
well as other implications on cosmological observables. Using this formalism,
we study the history of the thermalization process in high-resolution
hydrodynamic simulations of the Lambda cold dark matter model.  Collapsed
cosmological structures are surrounded by shocks with high Mach numbers up to
1000, but they play only a minor role in the energy balance of
thermalization. However, this finding has important consequences for our
understanding of the spatial distribution of CRs in the large-scale
structure. In high resolution simulations of galaxy clusters, we find a low
contribution of the averaged CR pressure, due to the small acceleration
efficiency of lower Mach numbers of flow shocks inside halos and the softer
adiabatic index of CRs. These effects disfavours CRs when a composite of
thermal gas and CRs is adiabatically compressed. However, within cool
core regions, the CR pressure reaches equipartition with the thermal pressure
leading there to a lower effective adiabatic index and thus to an enhanced
compressibility of the central intracluster medium. This effect increases the
central density and pressure of the cluster and thus the resulting X-ray
emission and the central Sunyaev-Zel'dovich flux decrement. The integrated
Sunyaev-Zel'dovich effect, however, is only slightly changed.

\section{Motivation}
Cosmological shock waves form abundantly in the course of structure formation,
both due to infalling cosmic plasma which accretes onto filaments, sheets and
halos, as well as due to supersonic flows associated with merging
substructures.  Additionally, shock waves in the interstellar and intracluster
media can be powered by non-gravitational energy sources, e.g.~as a result of
supernova explosions.  Cosmologically, shocks are important in several respects
for the thermal gas as well as for CR populations. (1) Shock waves dissipate
gravitational energy associated with hierarchical clustering into thermal
energy of the gas contained in dark matter halos, thus supplying the intra-halo
medium with entropy and thermal pressure support: where and when is the gas
heated to its present temperatures, and which shocks are mainly responsible for
it?  (2) Shocks also occur around moderately overdense filaments, heating the
intragalactic medium.  Sheets and filaments are predicted to host a warm-hot
intergalactic medium with temperatures in the range
$10^5\,\mbox{K}<T<10^7\,\mbox{K}$ whose evolution is primarily driven by shock
heating from gravitational perturbations developing into mildly nonlinear,
non-equilibrium structures. Thus, the shock-dissipated energy traces the large
scale structure and contains information about its dynamical history.  (3)
Besides thermalization, collisionless shocks are also able to accelerate ions
through diffusive shock acceleration. These energetic ions are reflected at
magnetic irregularities through magnetic resonances between the gyro-motion and
waves in the magnetized plasma and are able to gain energy in moving back and
forth through the shock front: what are the cosmological implications of such a
CR component, and does this influence the cosmic thermal history?  (4)
Simulating realistic CR distributions within galaxy clusters will provide
detailed predictions for the expected radio synchrotron and $\gamma$-ray
emission. What are the observational signatures of this radiation that is
predicted to be observed with the upcoming new generation of $\gamma$-ray
instruments and radio telescopes?

To date it is unknown how much pressure support is provided by CRs to the
thermal plasma of clusters of galaxies. A substantial CR pressure contribution
might have a major impact on the properties of the intracluster medium (ICM)
and potentially modify thermal cluster observables such as the X-ray emission
and the Sunyaev-Zel'dovich (SZ) effect. In contrast, CR protons play a decisive
role within the interstellar medium our own Galaxy.  CRs and magnetic fields
each contribute roughly as much energy and pressure to the galactic ISM as the
thermal gas does. CRs trace past energetic events such as supernovae, and they
reveal the underlying structure of the baryonic matter distribution through
their interactions.  CRs behave quite differently compared to the thermal
gas. Their equation of state is softer, they are able to propagate over
macroscopic distances, and their energy loss time-scales are typically larger
than the thermal ones.  Therefore, CR populations provide an important
reservoir for the energy from supernova explosions or structure formation shock
waves, and thereby help to maintain dynamical feedback for periods longer than
thermal gas physics alone would permit.

\section{Structure formation shock waves and cosmic rays}

\begin{figure}
  \begin{minipage}[t]{0.5\textwidth}
    \includegraphics[width=\textwidth]{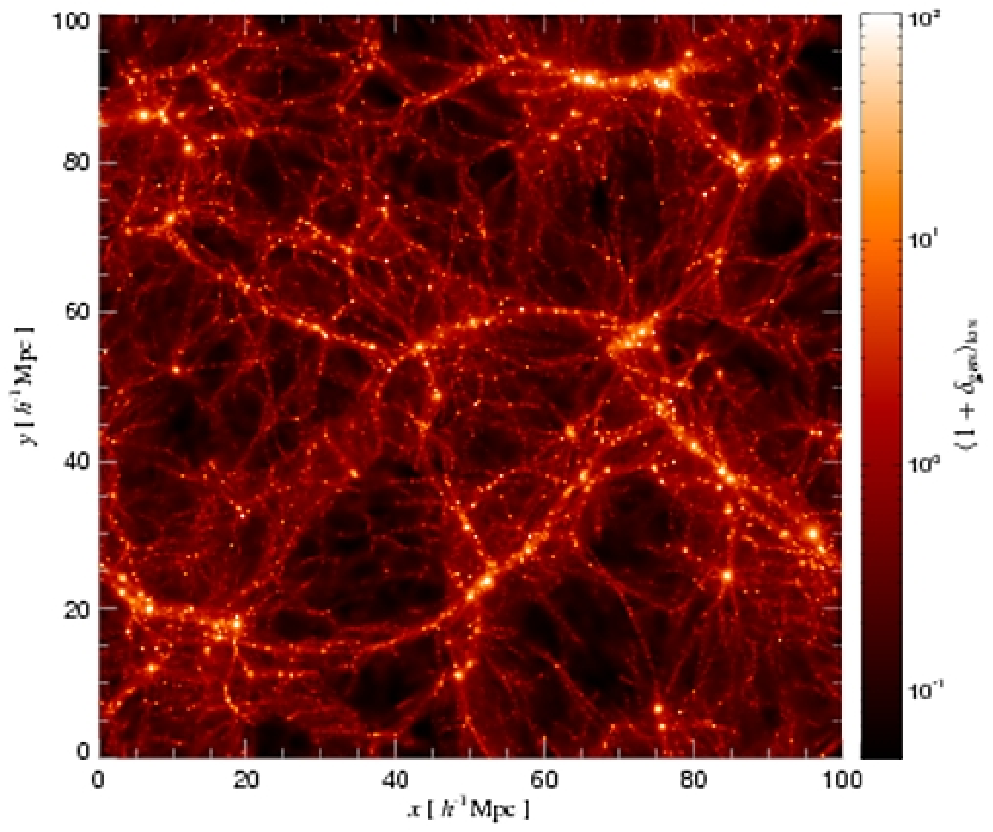}
    \includegraphics[width=\textwidth]{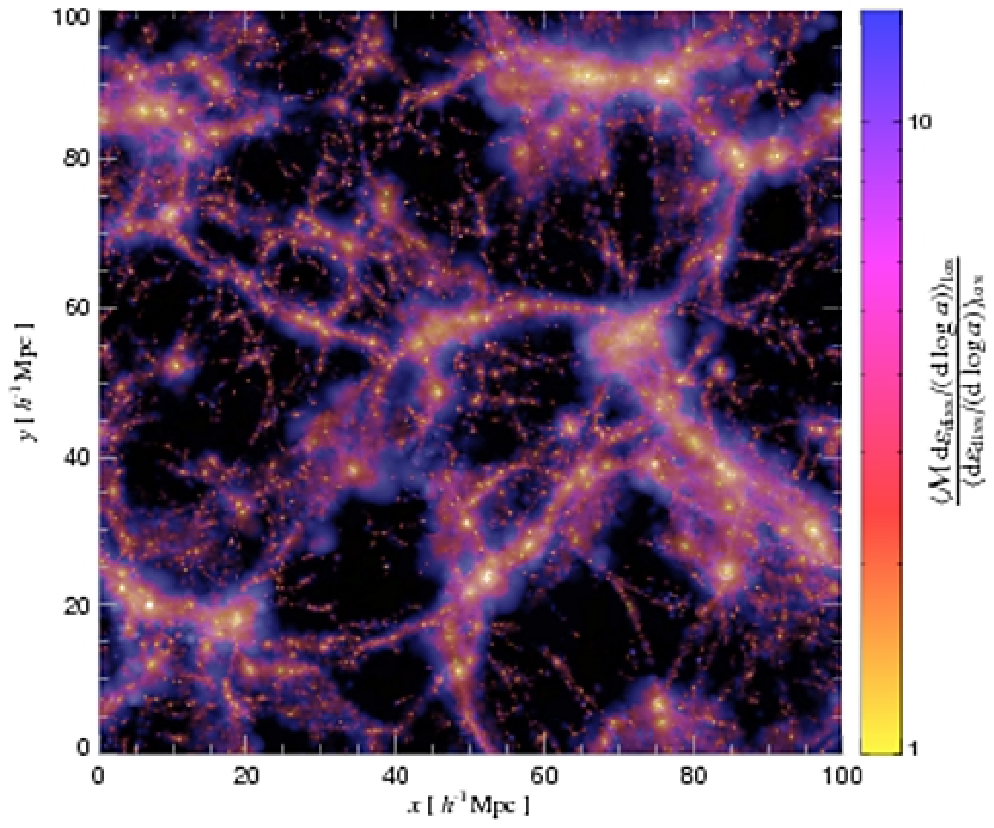}
  \end{minipage}
  \begin{minipage}[t]{0.5\textwidth}
    \includegraphics[width=\textwidth]{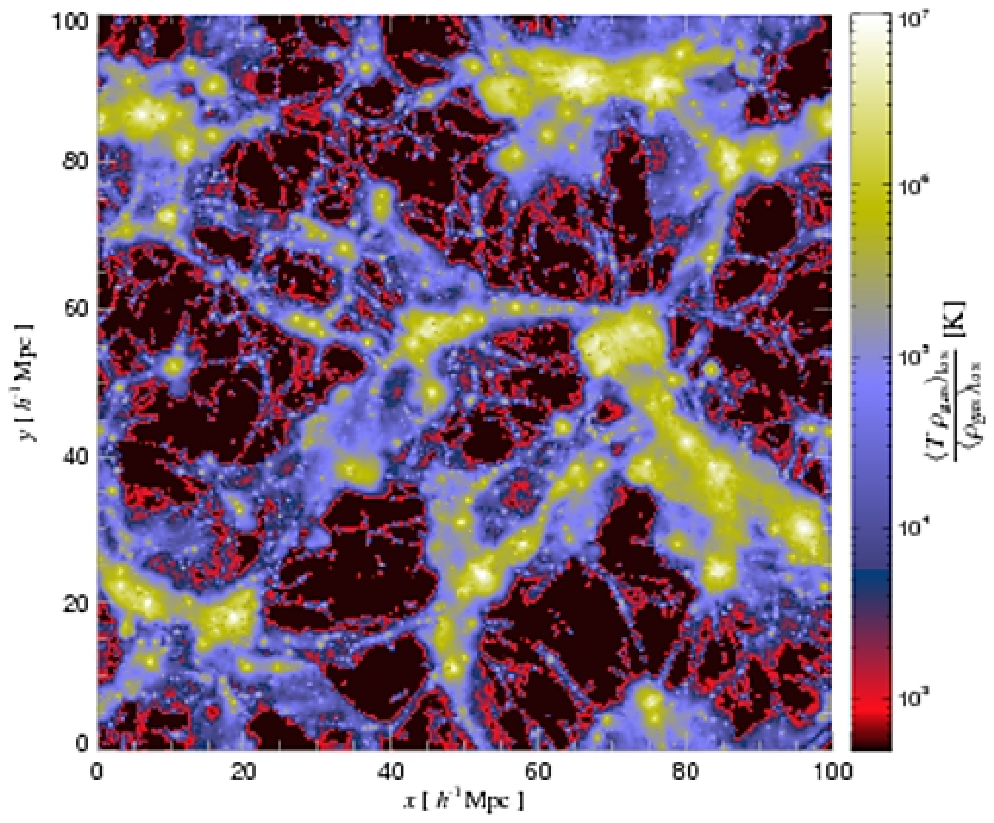}
    \includegraphics[width=\textwidth]{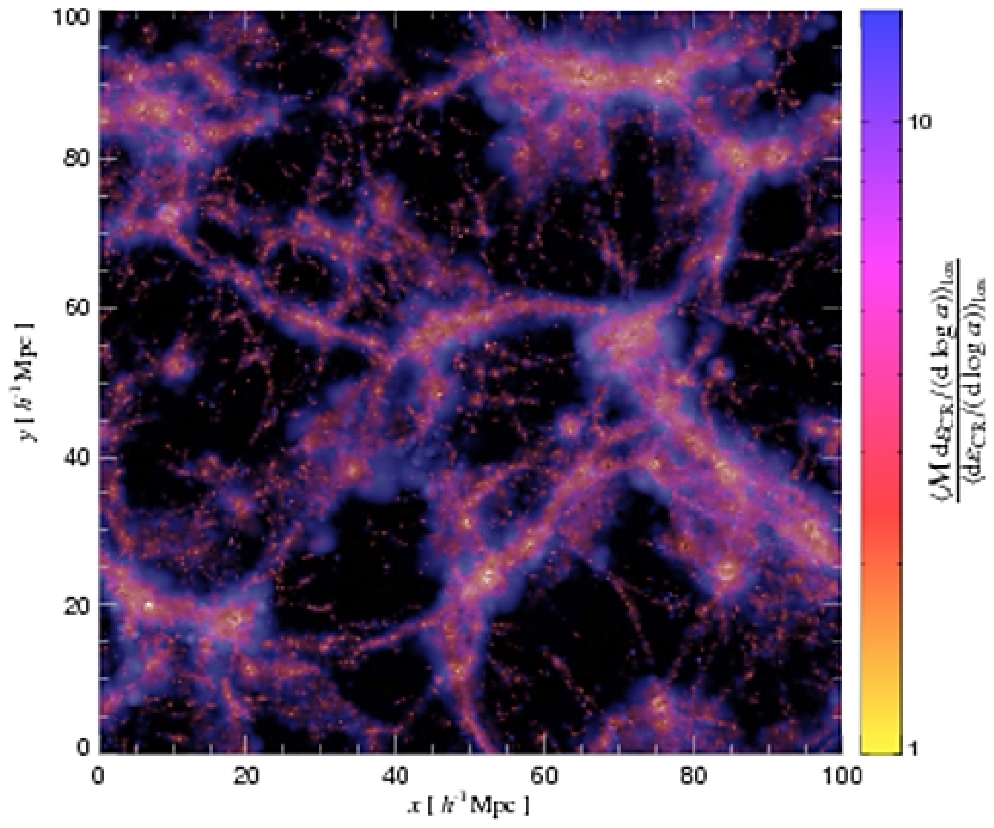}
  \end{minipage}
\caption{Visualization of a non-radiative cosmological simulation at redshift
  $z=0$ where the cosmic ray (CR) energy injection was only computed while the
  effect of the CR pressure on the dynamical evolution was not taken into
  account.  The {\em top panels} show the overdensity of the gas and the mass
  weighted temperature of the simulation. The {\em bottom panels} show a
  visualization of the strength of structure formation shocks. The colour hue of
  the map on the left-hand side encodes the spatial Mach number distribution
  weighted by the rate of energy dissipation at the shocks. The map on the
  right-hand side shows the Mach number distribution weighted by the rate of CR
  energy injection above the momentum threshold of hadronic CR p-p interactions.
  The brightness of each pixel is determined by the respective weights, i.e.~by
  the energy production density. Most of the energy is dissipated in weak
  shocks which are situated in the internal regions of groups or clusters,
  while collapsed cosmological structures are surrounded by strong external
  shocks (shown in blue). Since strong shocks are more efficient in
  accelerating CRs, the CR injection rate is more extended than the dissipation
  rate of thermal energy.}
\label{fig:cosmo}
\end{figure}

\begin{figure}
    \includegraphics[width=\textwidth]{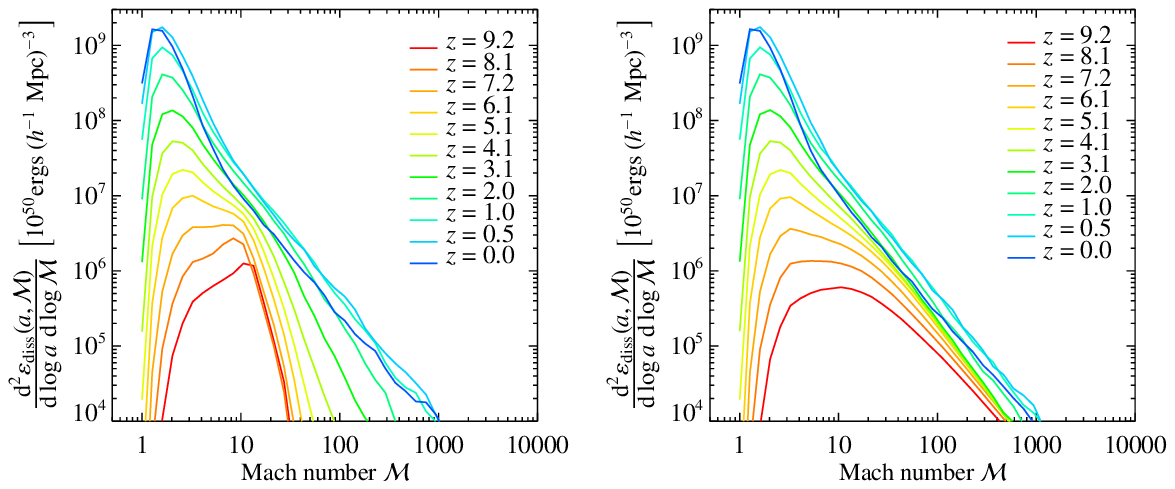}
\caption{Influence of reionisation (at redshift $z = 10$) on the Mach number
  statistics of non-radiative cosmological simulations. The figure on the {\em
  left-hand side} shows the differential Mach number distribution $\dd^2
  \eps_\rmn{diss}(a,\M)/( \dd \log a\, \dd \log\M)$ for our simulation with
  reionisation while the figure on the {\em right-hand side} shows this
  distribution for the simulation without reionisation.  Strong shocks are
  effectively suppressed due to an increase of the sound velocity after
  reionisation. }
\label{fig:distrib}
\end{figure}

We have developed a formalism that is able to measure the shock strength
instantaneously during an smoothed particle hydrodynamics (SPH) simulation
\cite{Pfrommer_a}. The method is applicable both to non-relativistic gas, and
to plasmas composed of CRs and thermal gas.  We apply our methods to study the
properties of structure formation shocks in high-resolution hydrodynamic
simulations of the Lambda cold dark matter ($\Lambda$CDM) model using an
extended version of the distributed-memory parallel TreeSPH code {\sc GADGET}-2
\cite{Springel} which includes self-consistent CR physics (\cite{Ensslin},
\cite{Jubelgas}). Fig.~\ref{fig:cosmo} shows the spatial distribution of
structure formation shocks in comparison to the density and temperature
distribution while Fig.~\ref{fig:distrib} shows the cosmological Mach number
distribution at different redshifts.\footnote{Note, that we corrected a missing
factor 10 in the normalization of Fig.6 in \cite{Pfrommer_a}.}

The main results are as follows. (1) Most of the energy is dissipated in weak
shocks internal to collapsed structures while collapsed cosmological structures
are surrounded by external shocks with much higher Mach numbers, up to $\M\sim
1000$. Although these external shocks play a major role locally, they
contribute only a small fraction to the global energy balance of
thermalization.  (2) More energy per logarithmic scale factor and volume is
dissipated at later times while the mean Mach number decreases with time. This
is because of the higher pre-shock gas densities within non-linear structures,
and the significant increase of the mean shock speed as the characteristic halo
mass grows with cosmic time.  (3) A reionisation epoch at $z_\rmn{reion}=10$
suppresses efficiently strong shocks at $z<z_\rmn{reion}$ due to the associated
increase of the sound speed after reionisation. (4) Strong accretion shocks
efficiently inject CRs at the cluster boundary.  This implies that the
dynamical importance of shock-injected CRs is comparatively large in the
low-density, peripheral halo regions, but is less important for the weaker flow
shocks occurring in central high-density regions of halos.

\section{Cosmic rays in hydrodynamic cluster simulations}

\begin{figure}
  \begin{minipage}[t]{0.5\textwidth}
    \centering{Relative CR pressure, \\non-radiative simulation:}
    \includegraphics[width=\textwidth]{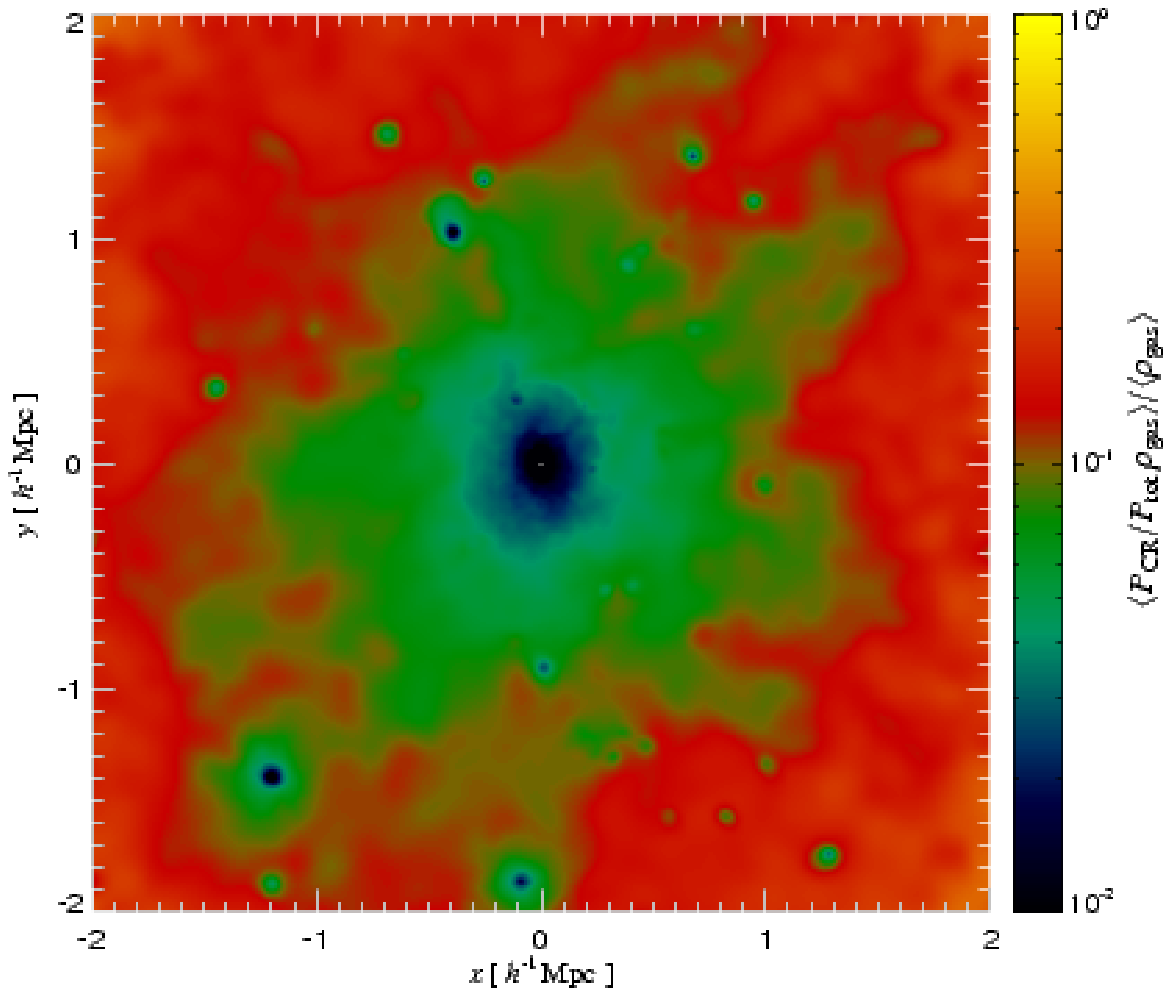}
    \centering{$S_X$ difference map:}
    \includegraphics[width=\textwidth]{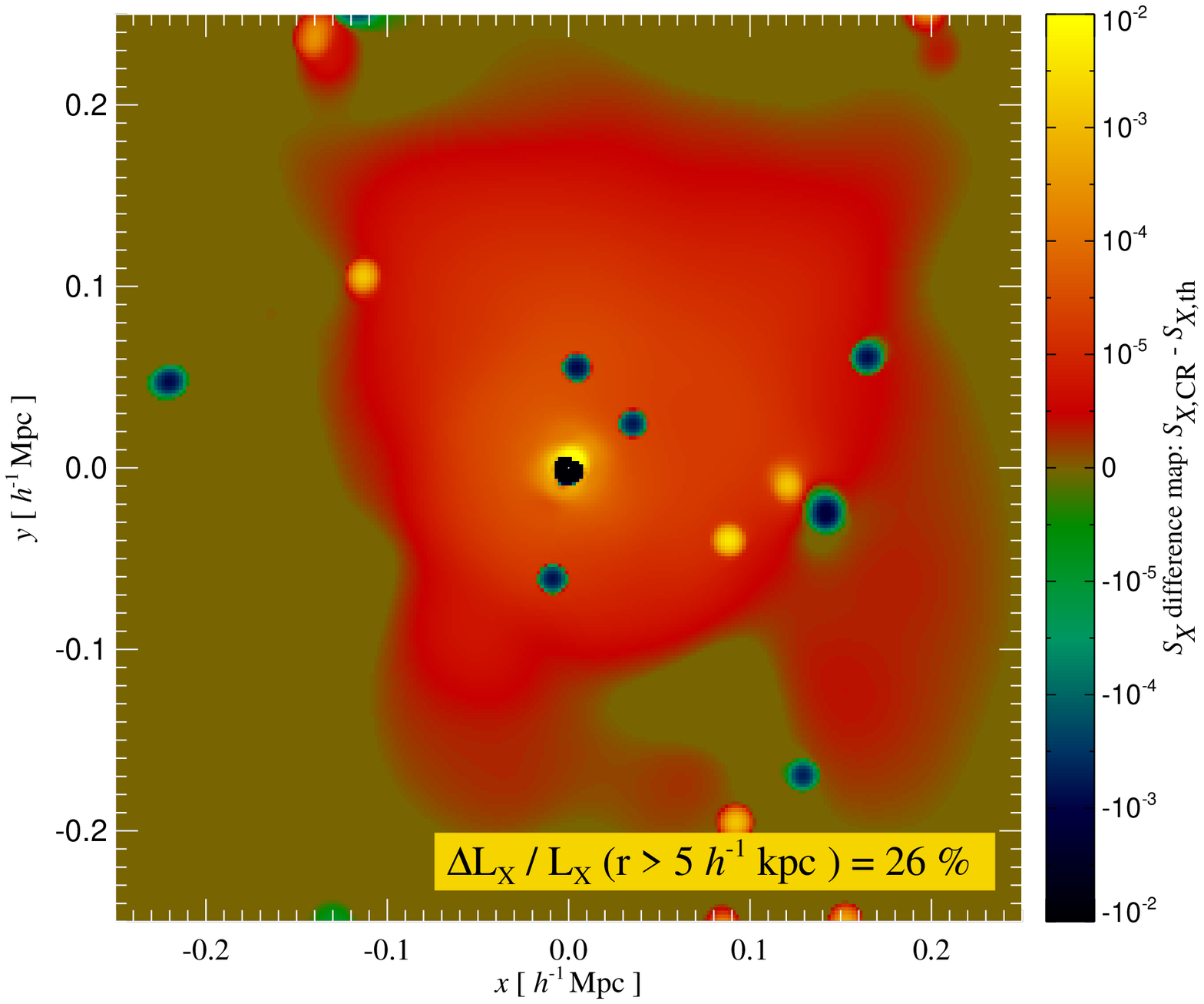}
  \end{minipage}
  \begin{minipage}[t]{0.5\textwidth}
    \centering{Relative CR pressure, \\radiative simulation:}
    \includegraphics[width=\textwidth]{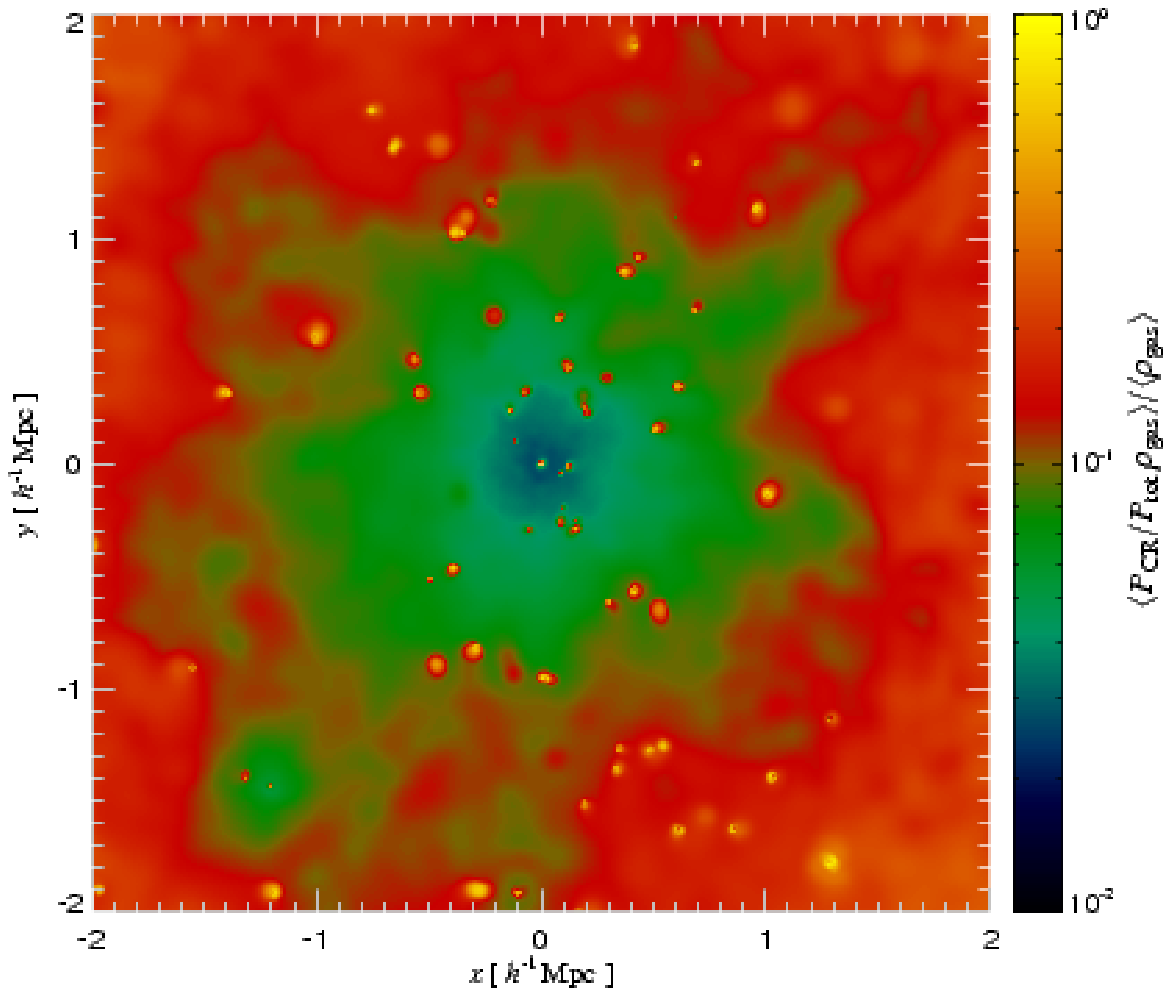}
    \centering{Compton-$y$ difference map:}
    \includegraphics[width=\textwidth]{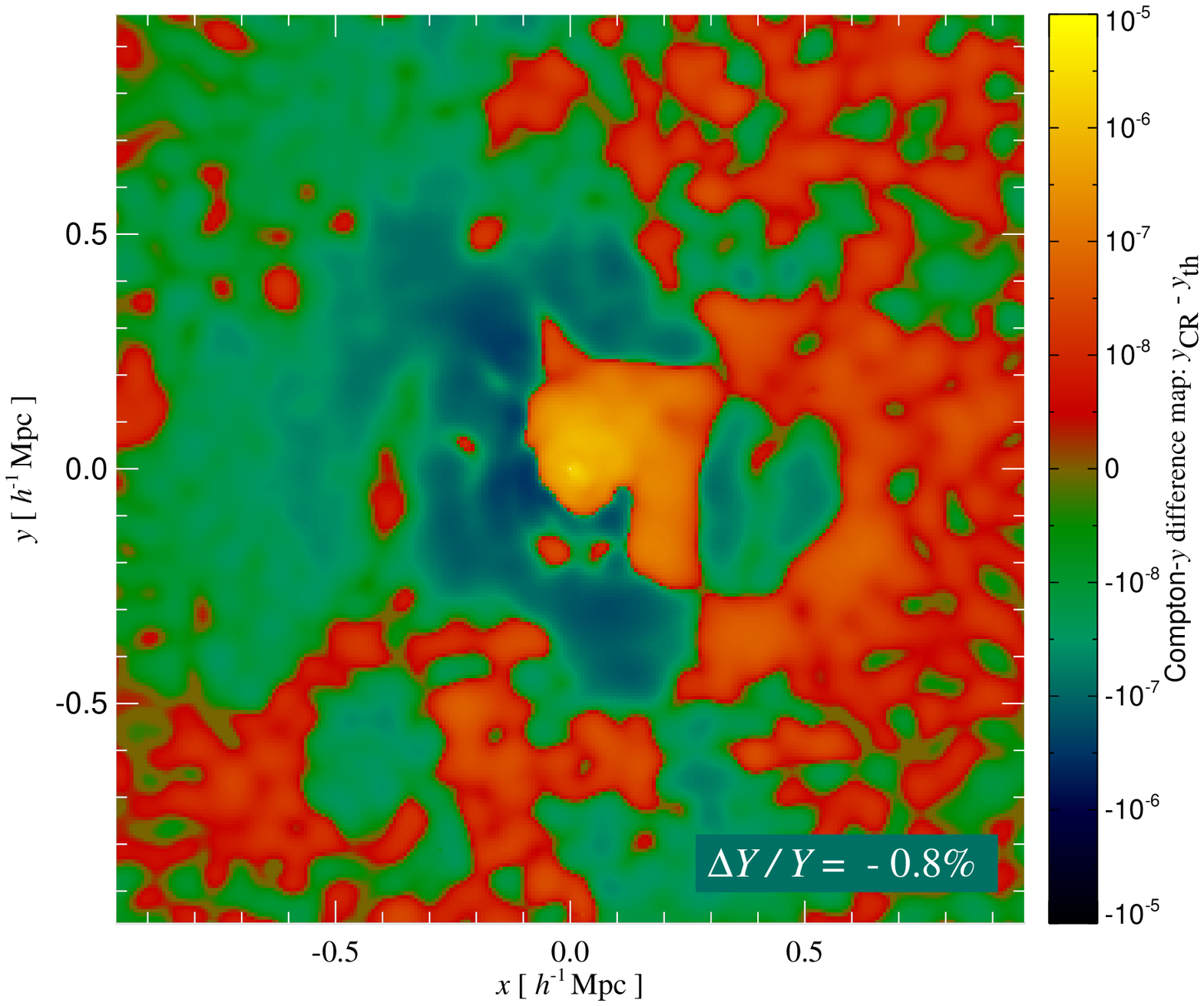}
  \end{minipage}
\caption{The top panels show a visualization of the pressure contained in CRs
  relative to the total pressure $X_\CR = P_\CR / (P_\CR + P_\therm)$ in a
  zoomed simulation of an individual galaxy cluster with mass $M = 10^{14}
  h^{-1} M_\odot$. The map on the {\em left-hand side} shows a non-radiative
  simulation with CRs accelerated at structure formation shock waves while the
  map on the {\em right-hand side} is from a simulation with dissipative gas
  physics including cooling, star formation, supernova feedback, and structure
  formation CRs. The lower panels show the CR-induced difference of the X-ray
  surface brightness $S_X$ ({\em left-hand side}) and the Compton-$y$ parameter
  ({\em right-hand side}) in a radiative simulation with structure formation
  CRs compared to the corresponding reference simulation without CRs. The
  relative difference of the integrated X-ray surface brightness/Compton-$y$
  parameter is given in the inlay. Within cool core regions, the CR pressure
  reaches equipartition with the thermal pressure, an effect that increases the
  compressibility of the central intracluster medium and thus the central
  density and pressure of the gas. This boosts the X-ray luminosity of the
  cluster and the central Sunyaev-Zel'dovich decrement while the integrated
  Sunyaev-Zel'dovich effect remains largely unaffected. }
\label{fig:XCR}
\end{figure}

To study the impact of CRs on cluster scales, we performed cosmological
high-resolution hydrodynamic simulations of a sample of galaxy clusters
spanning a large range in mass and dynamical states, with and without CR
physics. These clusters have originally been selected from a low-resolution
dark-matter-only simulation of a flat $\Lambda$CDM model and then re-simulated
using the 'zoomed initial conditions' technique.  We account for CR
acceleration at structure formation shocks and consider CR loss processes such
as their thermalization by Coulomb interactions and catastrophic losses by
hadronic interactions with ambient gas protons (see \cite{Pfrommer_b} for
details).  Within clusters, the relative CR pressure $X_\CR = P_\CR / (P_\CR +
P_\therm)$ declines towards a low central value of $X_\CR\simeq 10^{-4}$ in
non-radiative simulations due to a combination of the following effects: CR
acceleration is more efficient at the peripheral strong accretion shocks
compared to weak central flow shocks, adiabatic compression of a composite of
CRs and thermal gas disfavours the CR pressure relative to the thermal pressure
due to the softer equation of state of CRs, and CR loss processes are more
important at the dense centres. Interestingly, $X_\CR$ reaches high values at
the centre of the parent halo and each galactic substructure in our radiative
simulation due to the fast thermal cooling of gas which diminishes thermal
pressure support relative to that in CRs.  This additional CR pressure support
has important consequences for the thermal gas distribution at cluster centres
and alters the resulting X-ray emission and the SZ effect significantly
(cf. Fig.~\ref{fig:XCR}).

\section{Conclusions}

We studied the properties of cosmological shock waves using a technique that
allows us to identify and measure the shock strength on-the-fly during an SPH
simulation.  Invoking a model for CR acceleration in shock waves, we have
carried out the first hydrodynamical simulations that follows the CR physics
self-consistently. These simulations show that it is crucial to consider the
dynamical back-reaction of a non-thermal cosmic ray (CR) component in order to
describe the intracluster medium reliably. The X-ray luminosity from galaxy
clusters is boosted predominantly in low-mass cool core clusters due to the
large CR pressure contribution in the centre that leads to a higher
compressibility. The integrated Sunyaev-Zel'dovich effect is only slightly
changed while the central SZ flux decrement is also increased. These CR-induced
modifications can imprint themselves in changes of cluster scaling relations or
modify their intrinsic scatter and thus change the effective mass threshold of
X-ray or SZ surveys. Neglecting such a CR component in reference simulations
can introduce biases in the determination of cosmological parameters.

%
%

\printindex
\end{document}